\pdfoutput=1
\documentclass{JINST}
\usepackage{upgreek}
\usepackage{url}

\title{Development of Readout Interconnections for the Si-W Calorimeter of SiD}

\author{M.~Woods$^a$\thanks{Corresponding
author.}, R.~G.~Fields$^a$, B.~Holbrook$^a$, R.~L.~Lander$^a$, A.~Moskaleva$^a$, C.~Neher$^a$, J.~Pasner$^a$, M.~Tripathi$^a$, J.~E.~Brau$^b$, R.~E.~Frey$^b$, D.~Strom$^b$, M.~Breidenbach$^c$, D.~Freytag$^c$, G.~Haller$^c$, R.~Herbst$^c$, T.~Nelson$^c$, S.~Schier$^d$ and B.~Schumm$^d$\\
\llap{$^a$}University of California, Davis,\\
One Shields Ave, Davis CA 95616, USA\\
\llap{$^b$}University of Oregon,\\
  Eugene, OR 97403, USA\\
  \llap{$^c$}SLAC National Accelerator Laboratory,\\
  Menlo Park, CA 94025, USA\\
  \llap{$^d$}University of California, Santa Cruz\\
 1156 High Street, Santa Cruz, CA 95064, USA\\

  E-mail: \email{mwoods@ms.physics.ucdavis.edu}}

\abstract{The SiD collaboration is developing a Si-W sampling electromagnetic calorimeter, with anticipated application for the International Linear Collider. Assembling the modules for such a detector will involve special bonding technologies for the interconnections, especially for attaching a silicon detector wafer to a flex cable readout bus. We review the interconnect technologies involved, including oxidation removal processes, pad surface preparation, solder ball selection and placement, and bond quality assurance.  Our results show that solder ball bonding is a promising technique for the Si-W ECAL, and unresolved issues are being addressed.}


\begin{document}

\section{Introduction}
The International Linear Collider is a proposed e$^+$e$^-$ collider designed for precision physics at the energy frontier. The SiD concept, one of the two designs for particle detectors being designed and developed, includes a Si-W electromagnetic tracking calorimeter~\cite{brau}.  It features a thin onion-skin construction of detection (silicon) and interaction (tungsten) layers, allowing for implementation of particle flow algorithms. In order to achieve both a minimum transverse size of electromagnetic showers and a sufficient number of radiation lengths, while maintaining a reasonable size for the detector, thin layers are needed in the ECAL's structure. The detection layer, including sensor and readout hardware, is required to be no more than $\sim$1 mm thick~\cite{frey}.

\begin{figure}[tbp]
\begin{center}
\includegraphics[width=.7\textwidth]{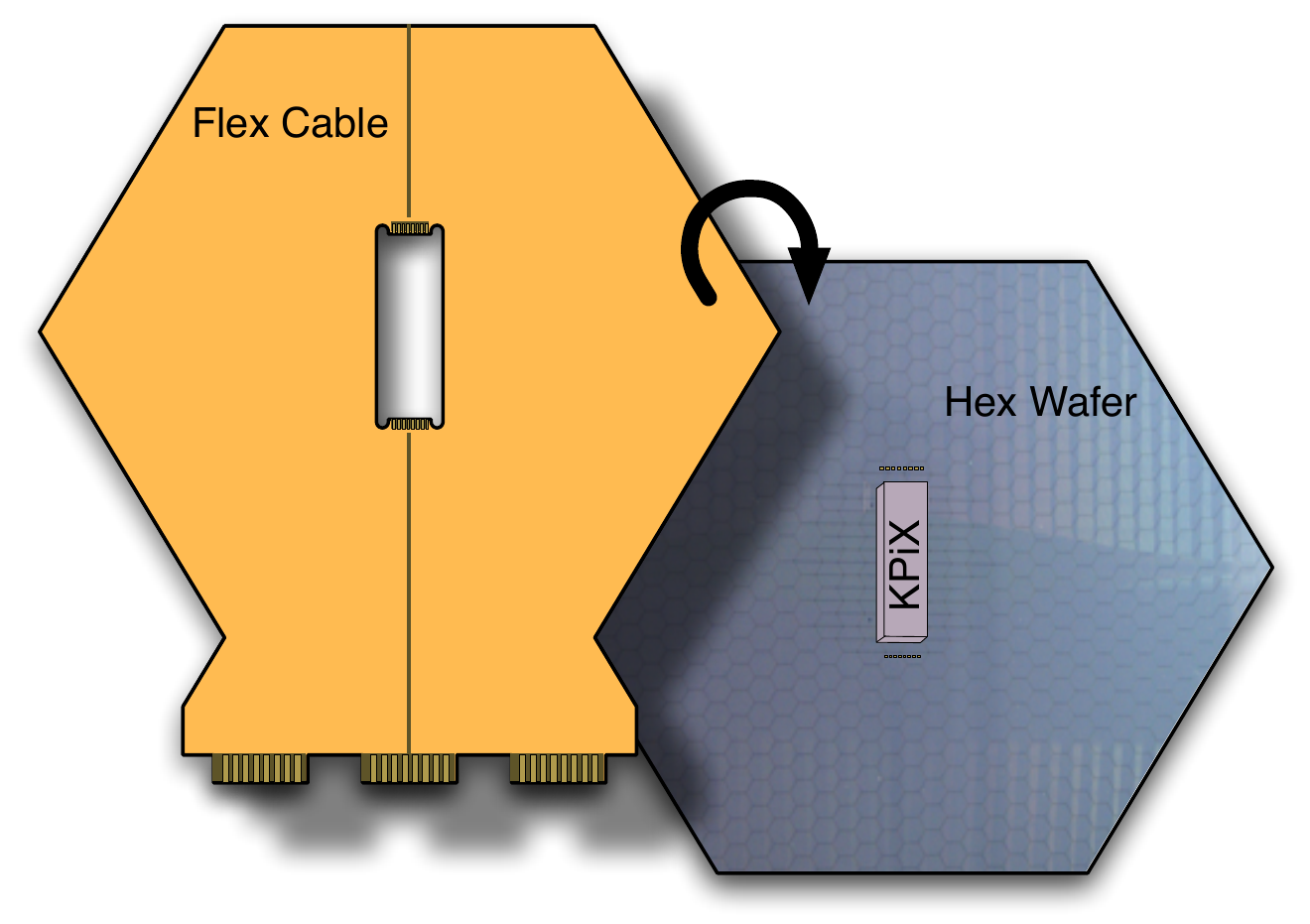}
\includegraphics[width=.2\textwidth]{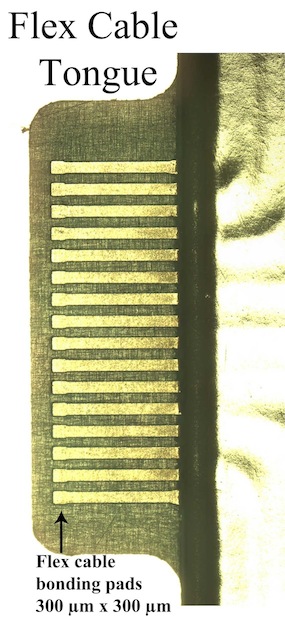}
\caption{An illustration of the assembly showing a flex cable before it is flipped and bonded to a hex wafer, which has a KPiX chip bonded on top (left).  The cut-out window in the cable has two ``tongues'' with pads, that extend out and are bonded to the data-bus pads on the sensor wafer. A closeup of the tongue of the flex cable (right) details the layout  of the traces coming from the kapton cable and the bond pads at the end of the traces. The bonding pads on the flex cable for flex cable to hex wafer attachment measure 300~$\upmu$m~$\times$~300~$\upmu$m with a 500~$\upmu$m pitch; the longest tongue dimension measures 10.3~mm.}
\label{fig:top-view}
\end{center}
\end{figure}

\begin{figure}[tbp]
\begin{center}
\includegraphics[width=0.95\textwidth]{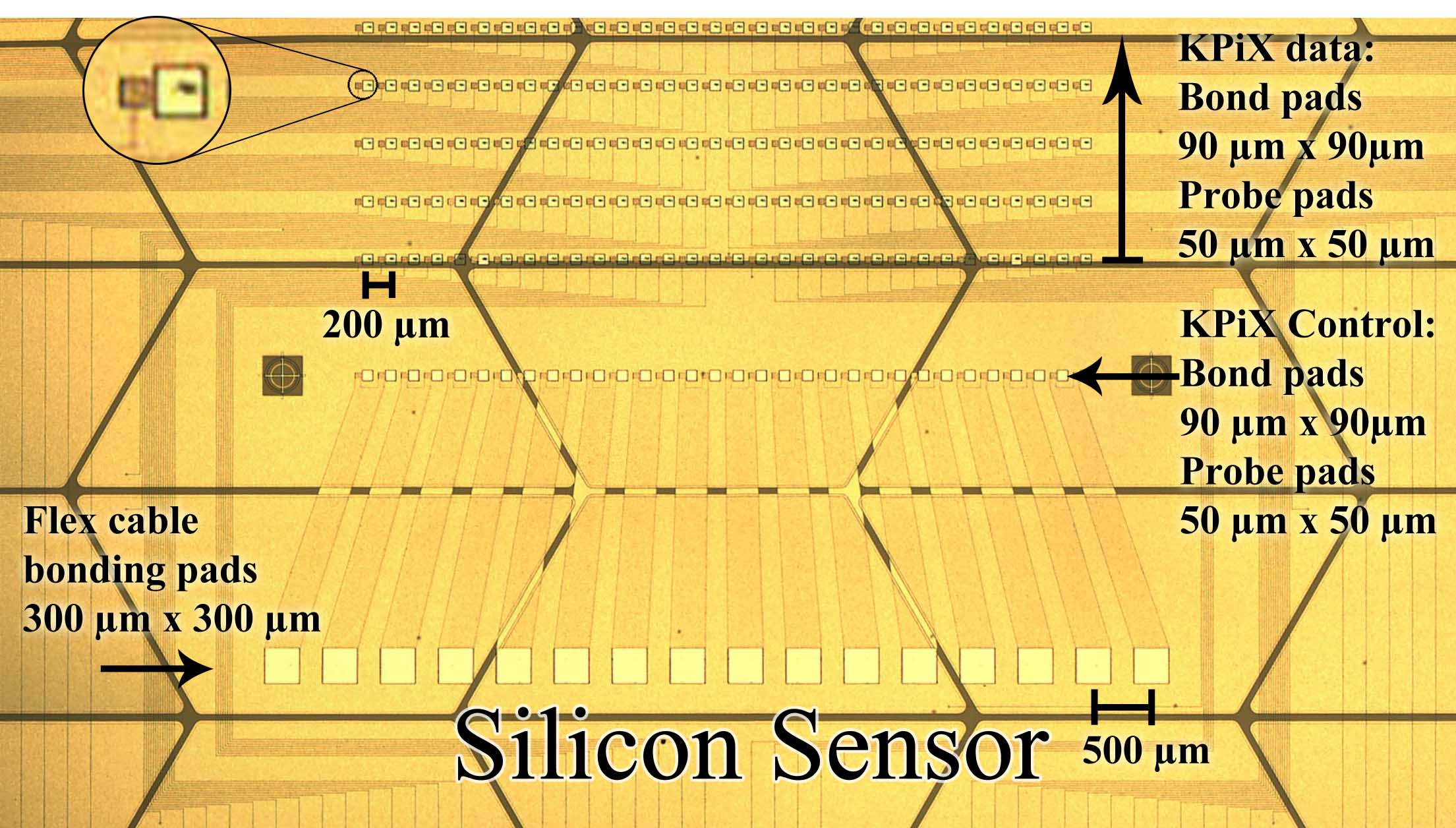}
\caption{A close-up photograph of the silicon hex wafer, illustrating the layout of the bonding parts. The 90~$\upmu$m~$\times$~90~$\upmu$m bonding pads in the central array are placed on a  200~$\upmu$m~$\times$~500~$\upmu$m grid. The pads for the flex cable attachment measure 300~$\upmu$m~$\times$~300~$\upmu$m on a linear array of 500~$\upmu$m pitch. Hex pixels and the web of 2~$\upmu$m pixel-to-pad interconnect traces are also visible.}
\label{fig:closeup-hex}
\end{center}
\end{figure}

The requirements of high granularity and thin readout layer are simultaneously met  by employing $\sim$6~inch hexagonal silicon sensors, which are segmented into $\sim$13~mm$^{2}$ pixels.  The sensors are read out by the KPiX chip, being developed at SLAC. The data readout bus consists of a hexagonally-stationed flexible kapton cable, which can handle up to sixteen KPiX chips. Currently, a version with a single station is used, as shown in Figure~\ref{fig:top-view}. Data transmission from the KPiX chip to the flex cable is accomplished via a set of short traces on the silicon sensor. Thus, the KPiX chip and the flex cable both need to be bonded to sets of pads on the silicon sensor. The bonding area on the flex cable, termed the ``tongue'', is shown in Figure~\ref{fig:top-view}. A closeup of the area of the silicon sensor that receives the flex cable is provided in Figure~\ref{fig:closeup-hex}.

Our previous work~\cite{tripathi} investigated the use of gold stud bonding as the interconnect technology, but it had to be abandoned because the oxide layer beneath the bonding pads on the sensor wafer was unable to withstand from the high pressure of thermo-compression bonding. Bump bonding using solder ball reflow offers the right solutions for bonding of both the KPiX and the flex cable to the sensor wafer. In this paper, we report on the development of the  solder bonding process for the Si-W ECAL. Most of the knowledge gained in this exercise is known to the interconnect industry but not widely disseminated, and had to be re-discovered for our custom application. At the UC Davis Facility for Interconnect Technology (UCD-FIT)~\cite{ucdfit}, this research has evolved not from the viewpoint of an industrial engineer, but with the approach of an experimental physicist faced with problems encountered in prototype development. Thus, we believe that the contents of this paper will provide insight to those developing similar platforms for future HEP applications.

\section{Bump Bonding Overview}

\begin{figure}
\begin{center}
\includegraphics[width=1\textwidth]{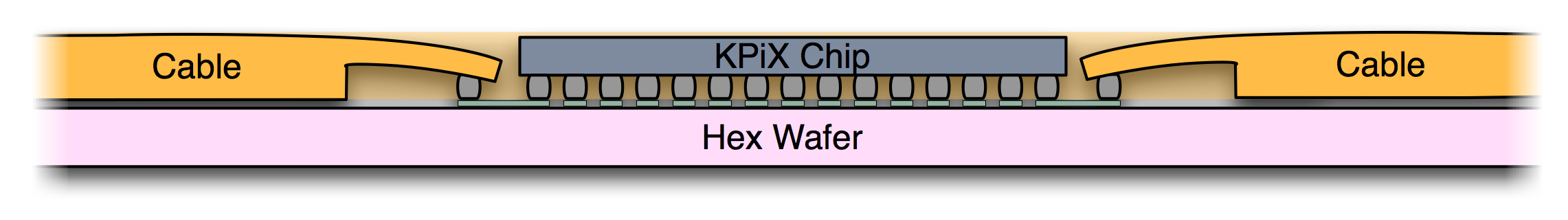}
\caption{Cross-section of a bonded single station of a single layer the Si-W ECAL with abbreviated number of pads. Solder balls (gray) placed on the KPiX chip and flex cable are bonded to the silicon sensor. Tongues extending from the flex cable make contact with exposed pads on top of traces embedded in the silicon sensor.}
\label{fig:cross-section}
\end{center}
\end{figure}

The Hamamatsu silicon detector has 1024 bonding pads that connect via buried metal traces to each of the 1024 pixel diodes on the sensor. The pads are to be bonded to matching pads on the KPiX chip, which form the inputs to individual channel amplifiers. An additional 64 pads on the KPiX, used to provide power, control signals and data output, are to be bonded to pads on traces leading to the flex cable. The KPiX pads are squares 70~$\upmu$m on a side and form a rectangular grid with a pitch of  200~$\upmu$m in the shorter dimension. Figure~\ref{fig:cross-section} shows a schematic representation of the flip chip bonding process required to establish mechanical and electrical connections and the layout of the pads and traces in the bonding region around the flex cable tongue. 

\subsection{The Flip-chip Process}

UCD-FIT houses a Finetech ``Fineplacer pico ma'' aligner-bonder which is a multipurpose die bonder with 5~$\upmu$m accuracy and repeatability~\cite{finetech}. A single camera views both the substrate (Si sensor) and package (KPiX) simultaneously using a rigid beam splitter. Engineered stability of the design guarantees that once the substrate material's position and angle visually overlaps that of the upper die, this orientation is maintained throughout the process of rotating the die with respect to the substrate and bonding them.  The other critical ability for the flip chip bonder is to ensure coplanarity of the KPiX as it is brought into contact with the sensor wafer. Even an angular discrepancy of only 0.1$^\circ$ across the length of the KPiX chip will create a height difference of 35~$\upmu$m. 

The solder  reflow process can be achieved in two ways: controlled gap and free-floating collapse.   In the former, maintaining a minimum gap between the upper die and the lower wafer can ensure a perfectly coplanar bonding scenario. Since the placement arm contains a heater (in addition to the bottom heater in contact with the wafer) and always maintains contact with the KPiX chip, both items are also heated equally during reflow. 

An alternative to the controlled gap is free floating solder bonding, in which the upper arm releases the die such that it rests on the lower wafer. This technique can be difficult with non-uniform  solder ball heights. Also, the KPiX chip is heated from only the bottom and the heat has to travel to the KPiX pads via the solder balls themselves. Any gap between solder and pad due to height difference prevents proper heating. An elongated heating time causes the solder to melt and allows its surface tension to pull the top chip into contact with more solder balls.  At this point, the new physical contacts start heating their corresponding bonding pads and the attachment proceeds. However,  it is greatly preferred to have pads at or near the temperature of the molten solder when contact is made. This may be achieved in a suitable reflow oven, but uniform heating of buried solder balls is a challenge due to the absence of convection. Hence, the preferred method for reflow is to use the controlled gap technique with uniform heating of the flip chip assembly.

Typically, a reducing chemical agent is used during bonding in order to remove oxides and/or prevent fresh oxidation. Pad preparation as a means to mitigate surface oxidation is discussed in the next section; here, we describe the use of liquid flux or forming gas to reduce the tin oxide formed in the Pb/Sn solder. Early on, we used a non-viscous liquid flux that was well suited for permeating the area around our solder balls via brush application. While this procedure showed success,  and using liquid flux became our benchmark process,  it was not desirable to have flux residue left over on the surface of sensitive chip electronics.

A  gas mixture of hydrogen (here, 5\%) and nitrogen (95\%) known as forming gas is capable of breaking down oxidation when the gas and samples are heated. Room temperature forming gas does little to prevent oxidation and its ability as a reducing agent is enhanced proportional to the temperature of the materials at hand. Forming gas is typically used in reflow ovens and can provide uniform oxidation removal without the need for a messy flux. We  outfitted our flip chip bonder with custom baffling which could support a forming gas environment around the parts being bonded. A study comparing the effectiveness of our liquid flux to the forming gas revealed that both were equally capable of removing and preventing oxide building up on the high temperature solder. Both methods resulted in multiple bonds with 100\% yield and resistances in the few m$\Omega$ range. Use of liquid flux was discontinued after confirming the forming gas system's effectiveness because it offers uniformity, repeatability and a residue-free finish that cannot be readily achieved with the manual application of liquid flux.

\subsection{Pad Surface Preparation}
The solder bonding process begins with the creation of appropriate bonding surfaces. The layers of materials used to build up bonding pads play a vital role in successful long term bonding for electrical components. Typically, silicon sensors and prototype ICs are fabricated with aluminum pads, which present a problem due to the presence of an insulating oxide layer which covers the entire aluminum surface.   In technologies like gold stud bonding, ultrasonic ball placement breaks through the thin (few nm) aluminum oxide layer, but for solder reflow processes the aluminum oxide layer must be addressed.

In the so-called electroless nickel immersion gold (ENIG) process, a zincate solution is used to remove aluminum oxide and deposit a zinc layer, which prepares the surface for electroless nickel plating. Immersion gold application follows to complete this process, providing a Au/Ni/Al metal stack. This is a preferred process for preparing solderable surfaces. The top layer of gold protects the nickel layer from oxidation and dissolves into molten solder during bonding; a strong solder bond is formed with the nickel layer.  The ENIG processing for our project was done at CVInc~\cite{CVI}.

In another technique, the metal stack can be built up by successive ion sputtering.  The wafer is coated with photo-resist and the windows are etched in the pad pattern.  First, reactive-ion etching (argon ions) is performed in order to remove about 10-20 nm of top layer of the pads.  This removes the oxide layer and some part of the underlying aluminum.  Next, a heavy metal like titanium, tungsten or platinum (or, their combination) is sputtered in order to form a barrier against migration of lighter ions placed above this layer.  This is followed by sputtering of nickel which forms the layer that is wettable by solder. Finally, a thin layer of gold is deposited in order to protect the nickel from oxidation.  The metal deposition for our dummy wafers (see below) was done in-house in the UCD micro fabrication facility. However, the processing of commercial chips was done at Advanced Research Corporation~\cite{ARC}, who have the capability to plasma etch the surface and sputter the ions without breaking the chamber vacuum. Alternately, a similar stack of metal layers can be deposited via electroplating.  This technique requires patterning and etch-back of metal layers.

\begin{figure}[tbp]
\begin{center}
\includegraphics[width=.55\textwidth]{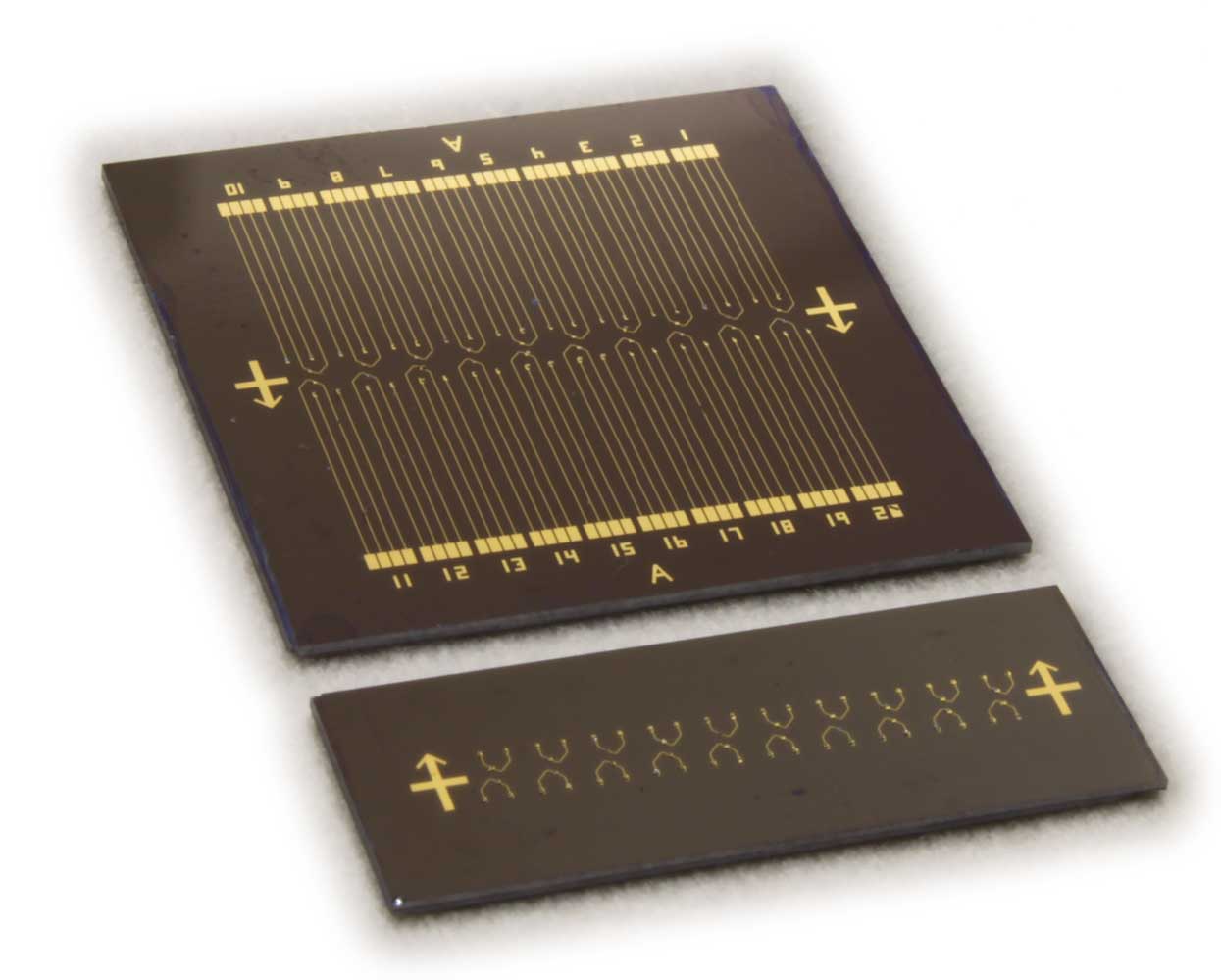}
\includegraphics[width=.354\textwidth]{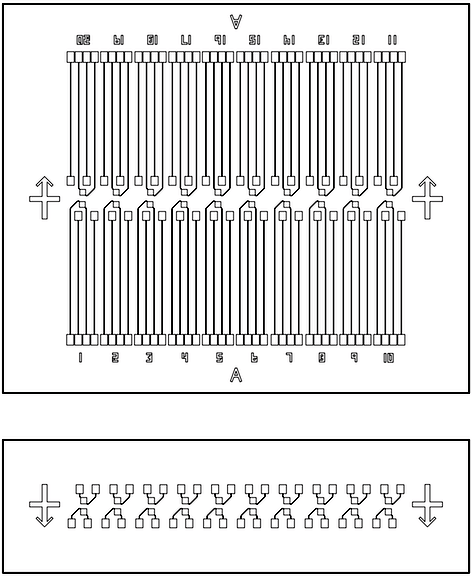}
\caption{(Left) A photograph of dummy chips (Au/Ni/Al/Cr), developed for characterizing solder bonding tests.  The smaller of the two chips is flipped and solder bonded on to the middle of the larger chip. (Right) A blow-up of the aluminum patterns. There are 20 sets of central pads for which 4-point resistance measurements of the bump bonds can be made.  There are 40 sets of auxiliary pads, also bump bonded, which are used for routing the traces to either end of the bump bonds. The numbers on the layout are enumerating each set of four measurement pads as they correspond to exactly one solder bond that can be measured for its resistance.}
\label{fig:topdown}
\end{center}
\end{figure}

\begin{figure}[tbp]
\begin{center}
\includegraphics[width=.45\textwidth]{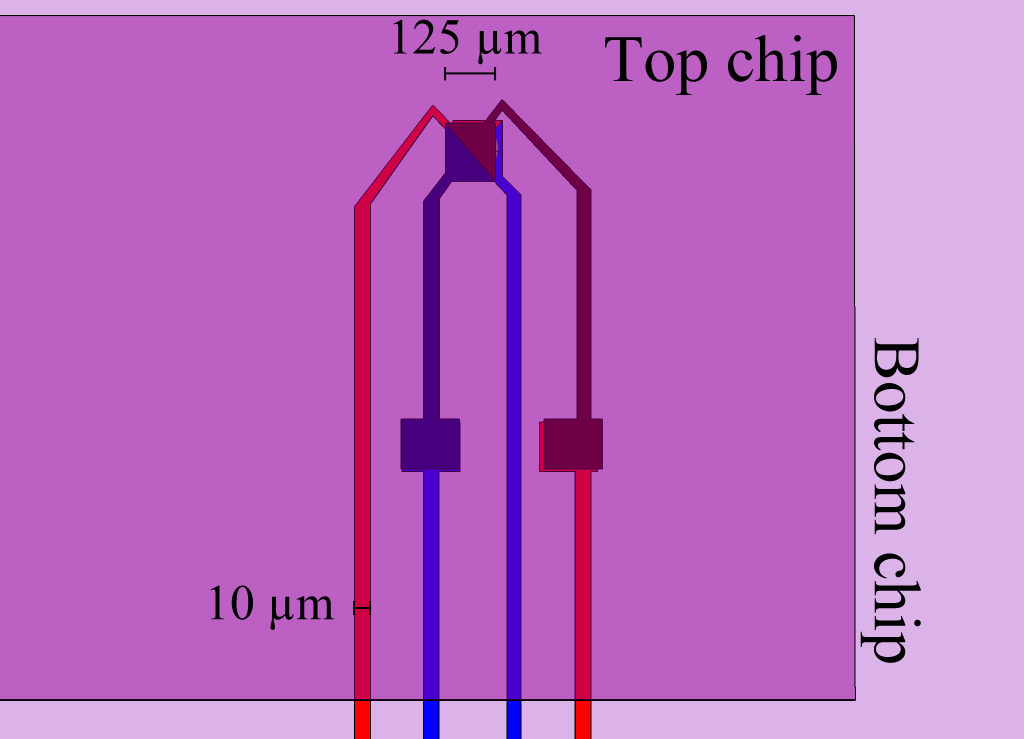}
\includegraphics[width=.45\textwidth]{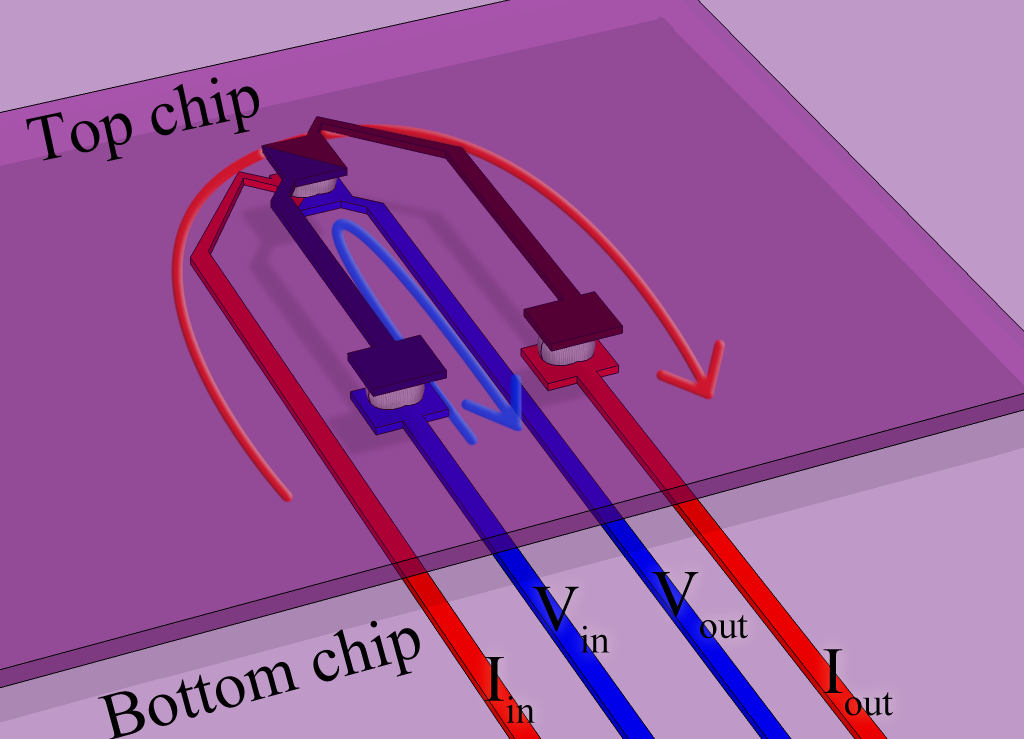}
\caption{Bonding orientation and resistance measurement procedure for one pad set (not to scale). The arrangement of the top (darker pads and traces) and bottom (lighter pads and traces) after bonding can be seen with a transparent top chip from directly over head (left). The three 125~$\upmu$m wide square bonding pads can be seen. The two on the lower portion serve only to allow current and voltage sensing to pass to the top chip. Only the resistance of the bond in the upper portion of the image is measured. A slant view (right) of the bonding scheme provides a glimpse of the bond and the (permutable) 4-point resistance measurement setup.}
\label{fig:dummy4point}
\end{center}
\end{figure}

\section{Studies of Solders with Dummy Chips}

The assembly sequence for a single station module consists of first bonding the KPiX chip to the wafer and then bonding the flex cable, in two separate bonding runs. The two stage nature necessitates the use of two separate solders with sufficiently distant transition temperatures such that the KPiX attachment is not threatened by the second stage bonding.  In order to study various solder types, we designed and fabricated dummy chips, as shown in Figure~\ref{fig:topdown}.  The bottom chip is larger with traces leading to a pattern of 60 pads, each 125~$\upmu$m on a side.  The top chip has a matching pad pattern with traces interconnecting them to allow for loop back resistance measurements.  This combination allows us to make 4-point resistance measurements of 20 bump-bonds, while the other 40 bonds, which transfer the resistance measurement's current and voltage sense to and from the upper chip's hidden traces, also contribute to yield measurements (Figure~\ref{fig:dummy4point}).

To address the concern of heating the Si-W assembly through two different solder reflow steps, a series of tests were done using dummy chips. Two eutectic solders were selected for testing: a high temperature (melting point of 183~$^\circ$C) eutectic lead/tin solder (Pb 63\%, Sn 37\%) for the KPiX bond and a low temperature solder (melting point of 143~$^\circ$C)  composed of indium/silver (In 97\%, Ag 3\%) for the flex cable.  The KPiX chips manufactured by the Taiwan Semiconductor Manufacturing Company (TSMC) come with the high temperature eutectic solder balls already placed on the pads.  For our studies, solder balls of the two types were placed on the dummy chips by CVInc.

\begin{figure}[tbp]
\begin{center}
\includegraphics[width=.45\textwidth]{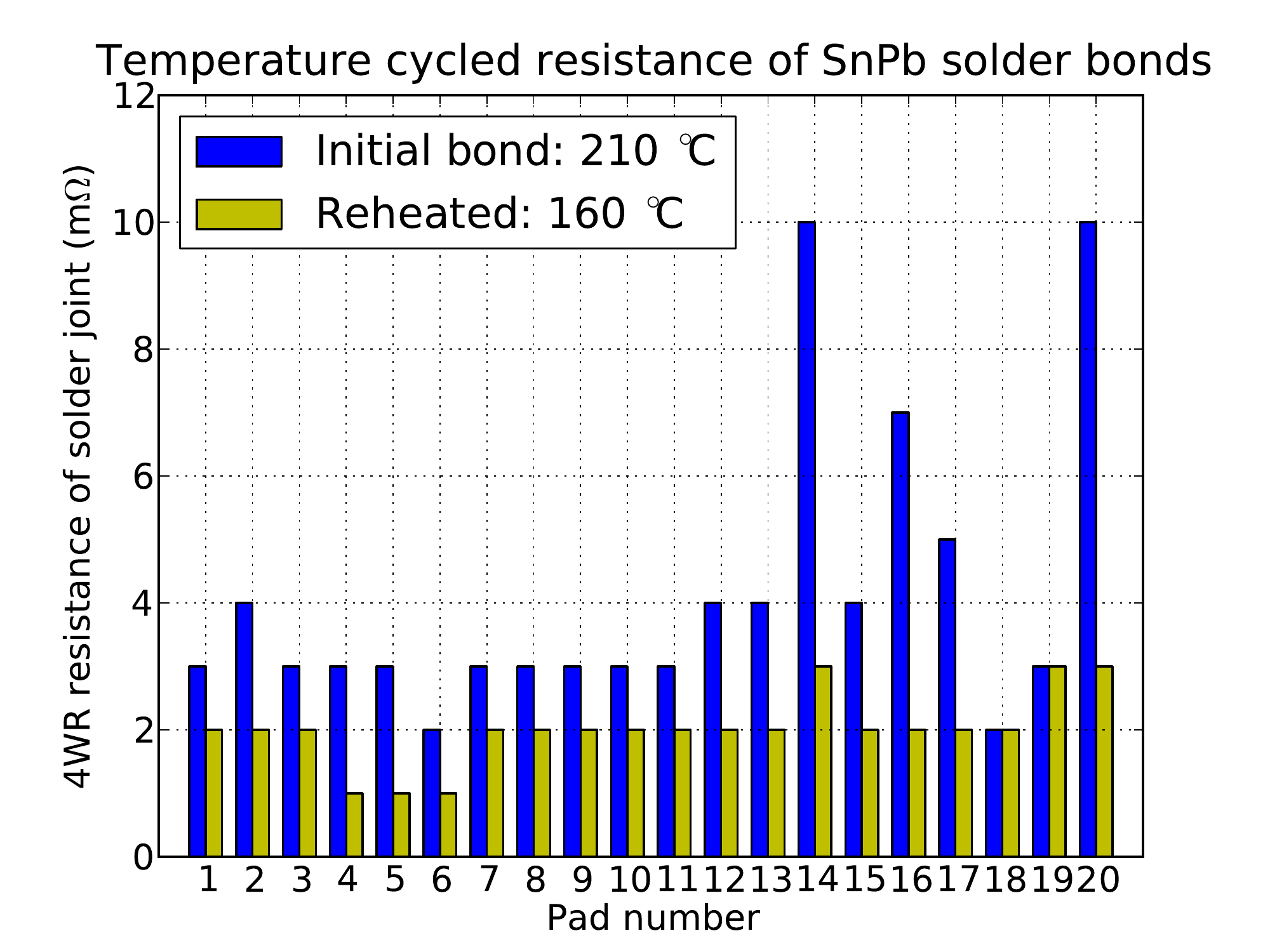}
\includegraphics[width=.45\textwidth]{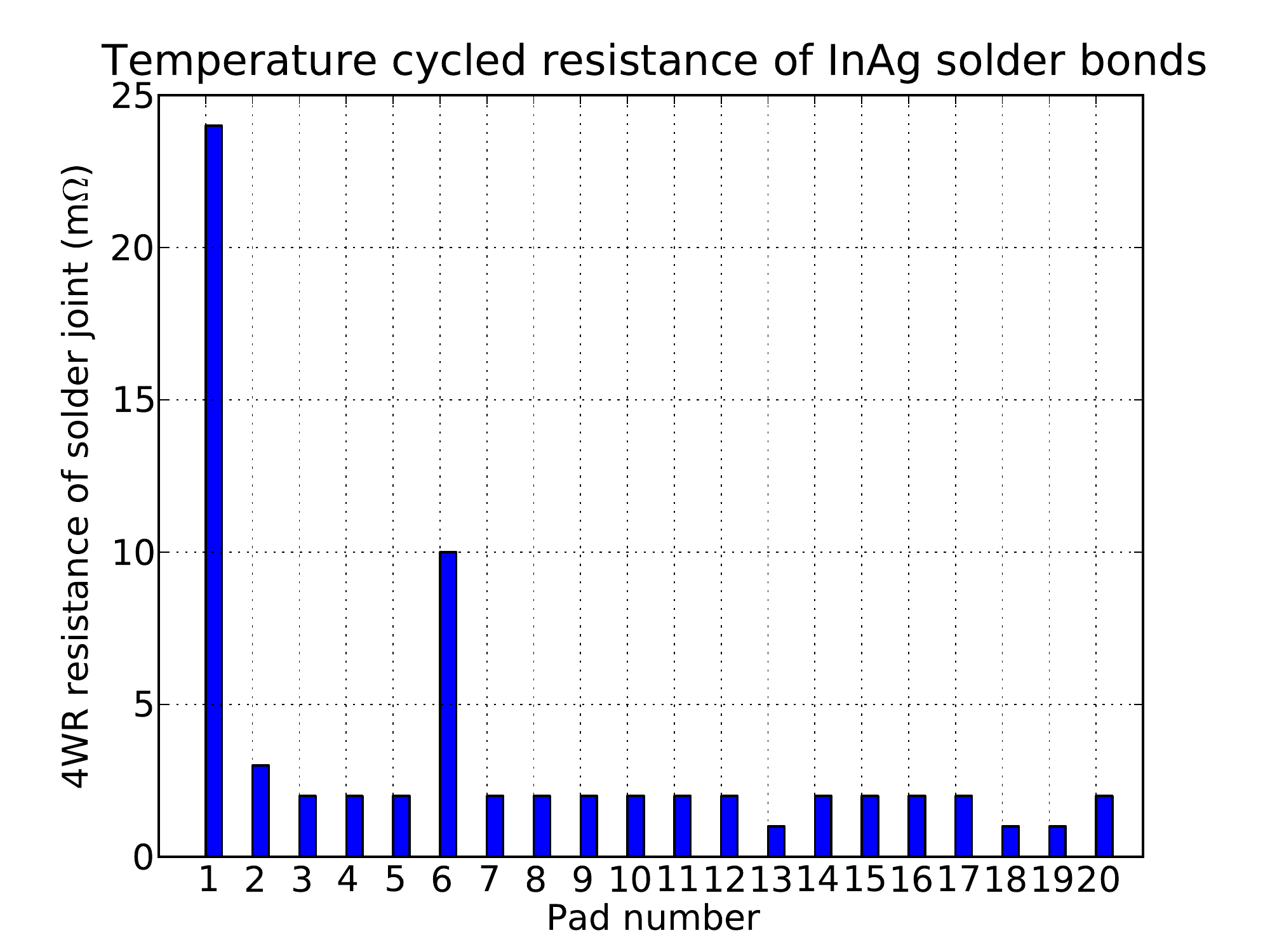}
\caption{Results from temperature cycling two kinds of solder. 
The left figure shows bond resistances of the high temperature SnPb solder (melting point 183 $^\circ$C) after bonding at 210 $^\circ$C, as well as the same solder recycled through the low temperature solder's bonding temperature. On the right is a plot showing that InAg low temperature solder (melting point 143 $^\circ$C) can be taken to 210 $^\circ$C before bonding and still produce resistances far under one ohm. Results are typical. The plot on the right is necessary only if the InAg solder is placed on the silicon sensor before bonding, not the flex cable. Both illustrate that no harm in cycling the solder is seen.}
\label{fig:dummyresults}
\end{center}
\end{figure}

There was worry that a high temperature solder joint, bonded in the first stage of an attachment, would weaken and/or fail when brought near its melting temperature during the second stage. To determine the effect of this temperature cycling, a pair of dummy chips was bonded using high temperature solder (at 210~$^\circ$C) and the resistances of the bond were measured. These values are shown in blue in Figure~\ref{fig:dummyresults}, left, where each of the twenty measured pads is shown. The maximum resistance seen is 10~m$\Omega$, which is  acceptable for the application.

The pair of bonded chips was then cycled up to 160~$^\circ$C, which is the temperature that is used for bonding the low temperature solder. Afterward, the same 4-point resistances were measured. The yellow bars in Figure~\ref{fig:dummyresults} show the resistance of the sample after reheating. The resistances have all decreased to a range of 1-3 m$\Omega$.  This lowering may be explained as due to domain walls of non-uniform alloys in the solder shifting and decreasing electrical resistance. Clearly the resistances did not \emph{increase} with the temperature cycle and thus this bonding scenario is acceptable.

If placed on the silicon sensor, the low temperature solder will also be reflowed during the bonding of the KPiX chip, but with nothing attached. Using dummy chips with low temperature solder, we first cycled the chip to a temperature of 210~$^\circ$C to imitate the high temperature solder bond; the chip was not being bonded at this high temperature, but rather simply exposed to the high temperature. Following this high temperature exposure, the sample was cooled and then bonded to its mating dummy chip at the usual 160~$^\circ$C. Figure~\ref{fig:dummyresults}, right, shows the 4-point resistance measurements following this bonding process. Each of the twenty pads is within an acceptable range. The highest value of 24 m$\Omega$ does not cause concern as it is still much smaller than one ohm, which is our acceptance level.

\section{Studies of Flex Cable Attachments}

\begin{figure}[tbp]
\begin{center}
\includegraphics[width=.35\textwidth]{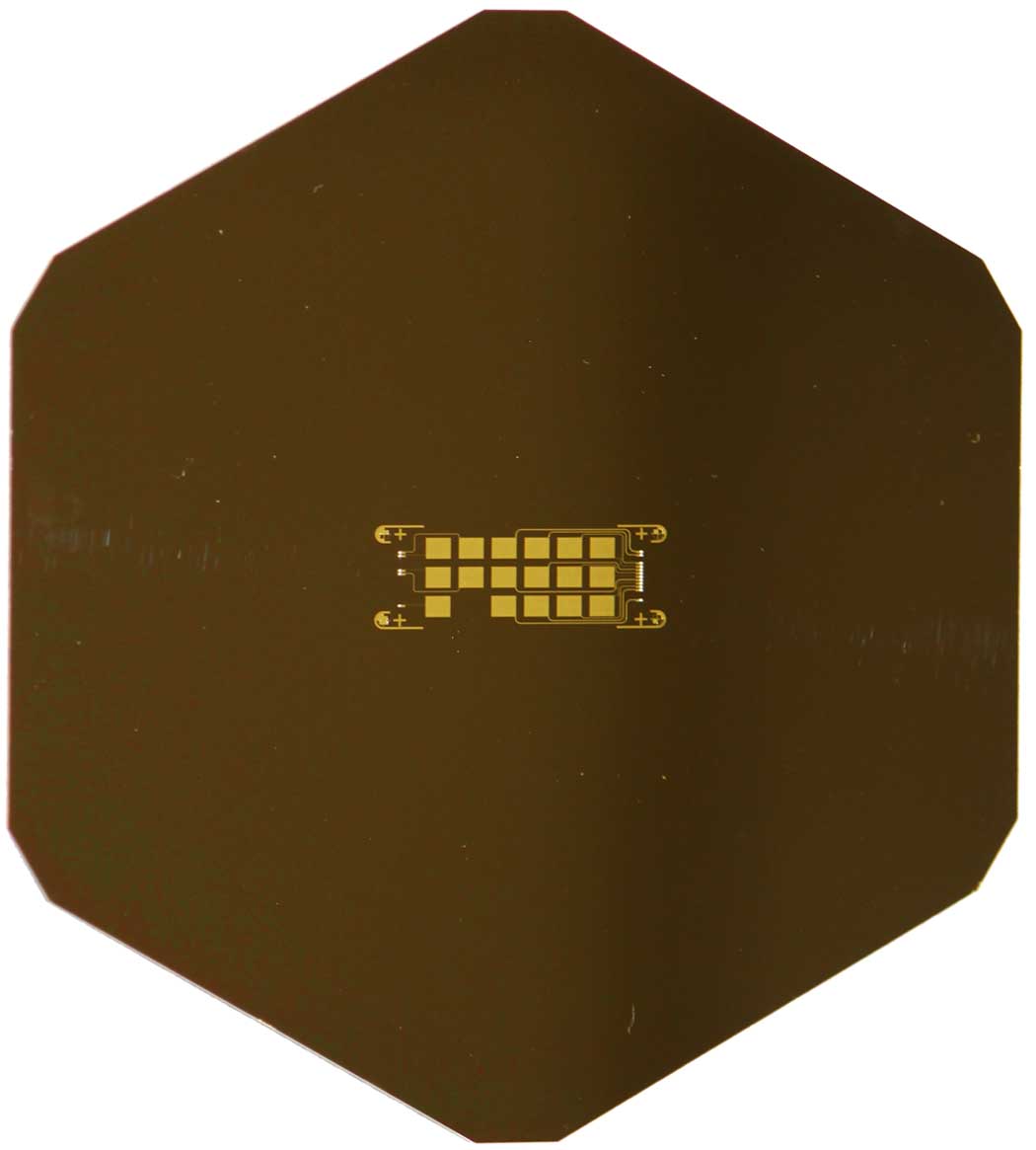}
\includegraphics[width=.51\textwidth]{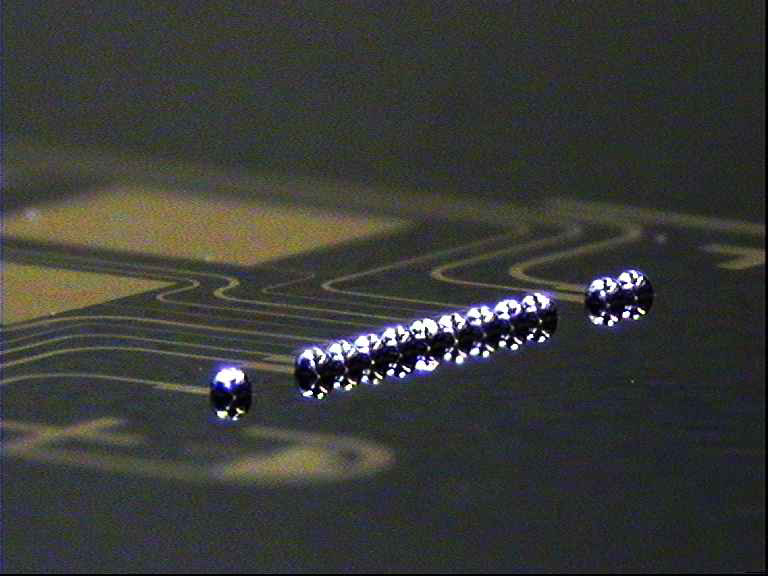}
\caption{Photographs of the dummy hex silicon wafer, which shares the same dimensions as the actual Hamamatsu sensors. Probe pads connecting to the cable bond pads are placed in the centre.  On the right is a close-up showing solder balls placed on pads that will receive the flex cable.}
\label{fig:dummyhex}
\end{center}
\end{figure}

After tests conducted using the dummy flip-chip components concluded that no negative effects were seen with the chosen types of solder, we initiated trials of flex cable attachment using the low temperature solder.   Because the silicon sensors are very expensive, we continued with use of dummy silicon sensors that allowed nearly identical bonding conditions without the financial risk. A new dummy hex wafer was fabricated to facilitate these studies, as shown in Figure~\ref{fig:dummyhex}.

Low temperature solder balls were placed by CVInc on the dummy sensor (and later the flex cable). After aligning the two components they were taken through the ``standard'' heating profile. The silicon dummy sensor and the kapton flex cable were brought up to a bonding temperature of 163 $^\circ$C over twenty seconds. The setup plateaus at this temperature for ten seconds before gradually cooling down over several minutes using forced air cooling.

It was found that good solder bonds were made between the pads of the dummy silicon sensor and one end of the flex cable. However, due to the difference in coefficients of thermal expansion (CTE) between the silicon and the kapton, the solder joints at the other end were ripped off upon cooling.  The CTE for the silicon hex wafer (or dummy chip as the case may be) is 2.6~ppm/$^\circ$C while the CTE for kapton polyimide is 20.0~ppm/$^\circ$C. The difference of 17.4~ppm/$^\circ$C in CTE across the 3~cm gap between the tongues of the flex cable for a temperature difference of 137~$^\circ$C seen during bonding produces a difference in expansion of 72~microns, a distance coincidentally close to the diameter of the solder balls.


\begin{figure}[tbp]
\begin{center}
\includegraphics[width=.8\textwidth]{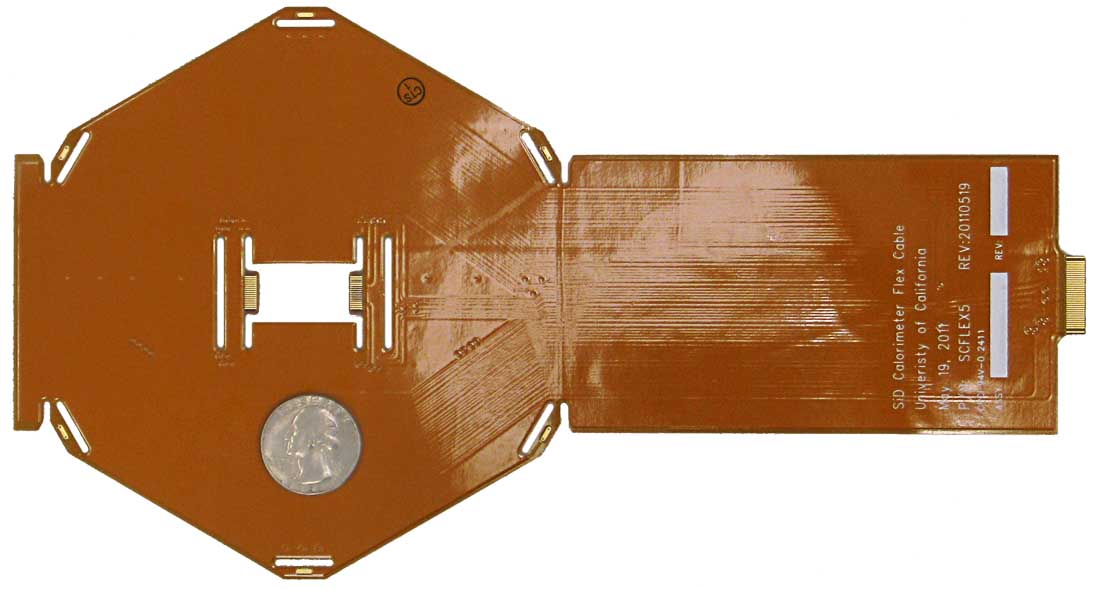}
\caption{A photograph of the flex cable showing strain relief slots added in critical areas. The slots near the center of the hexagonal portion of the flex cable help mitigate stress applied to the silicon sensor bonding pads. The slots near the edges of the flex cable are relieving the stress on guard ring pads. Texture due to buried trace routing is revealed in the reflection.}
\label{fig:slots}
\end{center}
\end{figure}

Strain relief slots were cut into the flex cable to ameliorate this problem, as shown in Figure~\ref{fig:slots}. The largest section of empty space between the two tongues is present to clear the KPiX chip, not to provide CTE mismatch relief. With slotting in place, temperature cycling could be performed without tearing traces off of silicon or shearing any solder away from bonding pads.

\begin{figure}[tbp]
\begin{center}
\includegraphics[width=.3\textwidth]{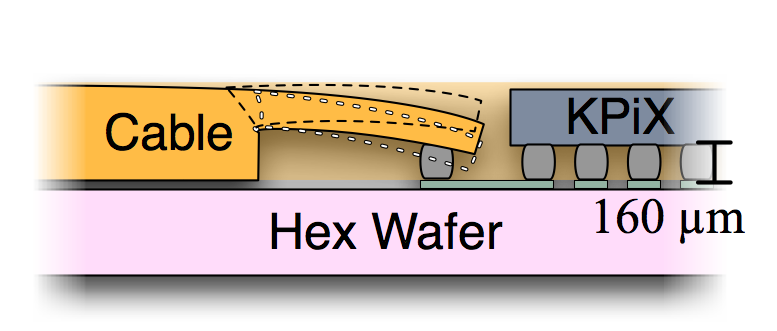}
\includegraphics[width=.65\textwidth]{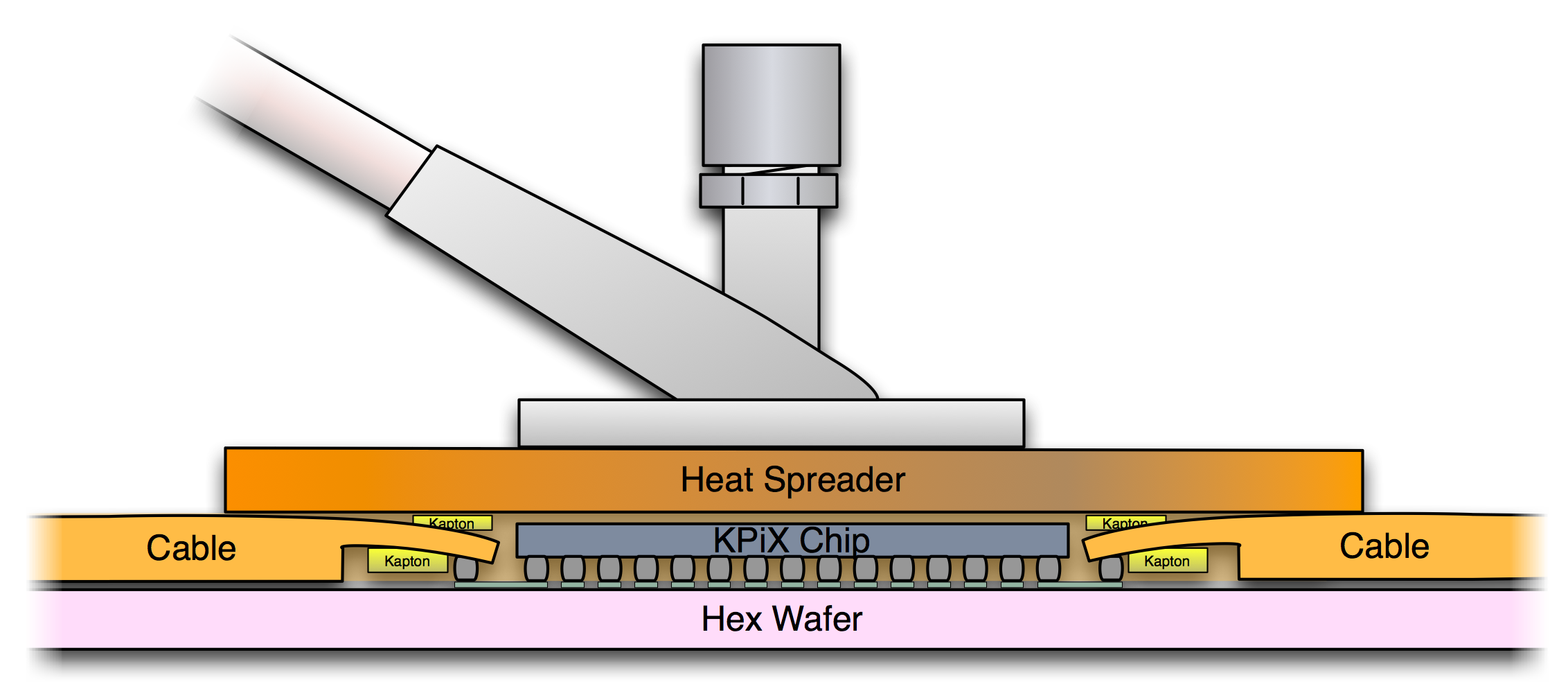}
\caption{Cross section of the flex cable bonding concerns and procedures. On the left, the dangers in variable kapton tongue bending are shown. The black and white outlines show upward tongue bending leaving a gap above the 160~$\upmu$m solder ball, creating an open bond and a low tongue that would significantly compress molten solder and risk solder bridging bonding sites and shorting pads together, respectively. On the right, the arm of the flip chip bonder bringing a heat spreader into contact with kapton tape above the tongues, pinching the tongues at an appropriate height with the surrounding cable thickness as a guide. }
\label{fig:kapton}
\end{center}
\end{figure}

Controlled gap reflow of the solder became an issue due to an unforeseen characteristic of the kapton tongues. Due to the flexible nature of the cable and internal stresses in the material left over from fabrication, the tongues see an unpredictable change in bend angle during heating. Steps were needed to ensure  that the bonding pad at the end of the tongue was present at the correct height above the hex wafer which was arbitrarily chosen as 60\% of the 160~$\upmu$m solder ball height; failure to control the bending could cause open circuits if the tongue was too high or short circuiting between bonds if the solder was violently compressed beyond individual pads and overflowing the solder dam structures in the cable if the tongue was too low (Figure~\ref{fig:kapton} (left)).

Compensation for the tongue bending required mechanical compensation to position the tongue. Because the conducting layer in the flex cable is located near the middle of the kapton cable's structure, surface material is removed to gain access to the conductive pads leading to a tongue that is thinner than the surrounding cable. We introduced thin strips of kapton tape underneath the tongue just next to the bonding pads and then above the tongue directly above the bonding pads as seen in Figure~\ref{fig:kapton} (right). The combined thickness of the two pieces of tape and the tongue was designed to equal the overall cable thickness. The kapton tape used was 4~mil thick for this solder ball. Only a single layer above and below was applied to provide sufficient stand-off. 

During bonding, a heat spreader that could encompass both tongues and wide enough to lay on the thicker areas of the flex cable was used to compress the assembly until the heat spreader was flush with the cable. The kapton standoffs corrected for excessive bending which allowed us to make viable solder bonds to the dummy hex wafer. With these issues under control, the flex cable to dummy hex wafer bond produced no open bonds, no shorts between pads, and reasonable sub-ohm resistance measurements.

 \section{Discussion}

Various solder reflow bump bonding tests have established this process as suitable for the interconnects required in the assembly of the Si-W calorimeter.  The use of two types of solders with different melting points has been proven as a successful method for sequential bump bonding on the same substrate.  An appropriate metal stack for solder bump bonding has been established. A flex cable designed with carefully placed strain-relief slots has been successfully bonded to silicon wafers without any detachment occurring upon cool down.  These results should provide important input to future designs involving interconnections between silicon chips and flex cables.

The preparation of pad surfaces for the hexagonal wafers produced by Hamamatsu~\cite{Hamamatsu} is proving to be a challenge.  They were treated by CVInc for the ENIG process but the zincating process failed to produce a plating layer.  An attempt to build up a metal stack of Au/Ni/Ti via a sputtering process was attempted by ARC.  However, the stack detached during lift-off, leaving the original surface behind.  We are investigating the surface composition of the pads in order to understand the cause, while also pursuing a plating and etch-back technique with Fraunhofer IZM~\cite{IZM}. We do not understand whether there are deficiencies in the metallization techniques or whether the pad surface of the Hamamatsu detectors is anomalous.  Flip-chip bonding of ASICs to commercial aluminum substrate pads is a well-developed process. We can expect eventually to find the source of our problem, be it the Hamamatsu pads or our processing, and make successful bonds of the ASIC to the sensor wafer.  If the problem is the former, we will need to have Hamamatsu fabricate new wafers with gold pads.  In summary, while there are remaining issues with sensor pad surface, the interconnect issues have all been solved.
 
\acknowledgments

        This work is funded by the Linear Collider Detector R\&D Program, sponsored by the DOE Office of Science. We thank Steve Ellison (ARC) and Terence Collier (CVInc) for numerous technical discussions.


\begin{thebibliography}{9}

\bibitem{brau} J. Brau et al., \emph{An electromagnetic calorimeter for the SiD concept}, 2007 \emph{Pramana}, Vol. 69, No.6, pp. 1025-30.
 
\bibitem{frey}R. Frey et al., \emph{A Silicon-Tungsten ECal with Integrated Electronics}, 2007, arXiv:0710.2373v1 [physics.ins-det]

\bibitem{tripathi}M. Tripathi et al., \emph{Gold-stud Bump Bonding for HEP Applications}, \jinst{5}{2010}{C08005}, \url{http://iopscience.iop.org/1748-0221/5/08/C08005/}

\bibitem{ucdfit} UC Davis Facility for Interconnect Technology, \url{http://fit.physics.ucdavis.edu/}

\bibitem{finetech} Finetech USA, \url{http://www.finetechusa.com/}

\bibitem{CVI} CVInc,  \url{http://www.covinc.com/}

\bibitem{ARC} Advanced Research Corporation,  \url{http://www.arcnano.com/}

\bibitem{Hamamatsu} Hamamatsu Photonics, K. K., \url{http://www.hamamatsu.com/}

\bibitem{IZM} Fraunhofer IZM, \url{http://www.izm.fraunhofer.de/}


\end{thebibliography}
\end{document}